\documentclass[prd,preprint,showpacs]{revtex4}
\usepackage{graphicx}

\usepackage{amssymb}

\def\be{\begin{equation}}
\def\ee{\end{equation}}
\def\bea{\begin{eqnarray}}
\def\eea{\end{eqnarray}}

\def\d{\rm{d}}

\def\S{\rm{S}}
\def\vr{\varrho}
\newcommand{\stg}{{\sqrt{-\tilde{g}}}}
\newcommand{\sg} {{\sqrt{-g}}}

\newcommand{\bi} {\begin{itemize}}
\newcommand{\ei} {\end{itemize}}
\newcommand{\ben} {\begin{enumerate}}
\newcommand{\een} {\end{enumerate}}

\newcommand{\LL} {{\cal L}}

\newcommand{\del} {\partial}
\newcommand{\delsl} {\partial\hspace{-1.2ex}/}

\newcommand{\rar} {{\rightarrow}}

\newcommand{\la} {{\lambda}}
\newcommand{\al} {{\alpha}}

\newcommand{\nl} {{\newline}}
\newcommand{\Om} {{\Omega}}
\newcommand{\om} {{\omega}}
\newcommand{\de} {{\delta}}

\newcommand{\si} {{\sigma}}

\newcommand{\Ga} {{\Gamma}}
\newcommand{\ga} {{\gamma}}

\def\v1{\vspace{1cm}}

\begin{document}
\title{Conformal transformations and conformal invariance in gravitation}

\author{Mariusz P. D\c{a}browski}
\email{mpdabfz@wmf.univ.szczecin.pl}
\author{Janusz Garecki}
\email{garecki@wmf.univ.szczecin.pl}
\affiliation{\it Institute of Physics, University of Szczecin, Wielkopolska 15,
          70-451 Szczecin, Poland}
\author{David B. Blaschke}
\email{blaschke@ift.uni.wroc.pl}
\affiliation{Institute for Theoretical Physics, University of Wroc{\l }aw,
Max Born Pl. 9, 50-205 Wroc{\l }aw, Poland}
\affiliation{Bogoliubov  Laboratory of Theoretical Physics, Joint
Institute for Nuclear Research,
Joliot-Curie Street 6, 141980  Dubna, Russia}
\date{\today}

\begin{abstract}
Conformal transformations are frequently used tools in order to study relations between
various theories of gravity and the Einstein relativity. In this paper
we discuss the rules of these transformations for geometric quantities
as well as for the matter energy-momentum tensor. We show 
the subtlety of the matter energy-momentum conservation law which refers to the fact that
the conformal transformation ``creates'' an extra matter term composed
of the conformal factor which enters the conservation law. In an
extreme case of the flat original spacetime the matter is
``created'' due to work done by the conformal transformation to
bend the spacetime which was originally flat.
We discuss how to construct the conformally invariant gravity
theories and also find the conformal transformation rules for the
curvature invariants $R^2$, $R_{ab}R^{ab}$, $R_{abcd}R^{abcd}$ and 
the Gauss-Bonnet invariant in a spacetime of an arbitrary dimension.
Finally, we present the conformal transformation rules 
in the fashion of the duality transformations of the superstring theory. In such
a case the transitions between conformal frames reduce to 
a simple change of the sign of a redefined conformal factor.

\end{abstract}


\maketitle


\section{Introduction}
\label{sect1}

\setcounter{equation}{0}

Conformal transformations of the metric tensor \cite{hawk_ellis} are interesting
characteristics of the scalar-tensor theories of gravity
\cite{bd,maeda,scaltens,polarski}, including its conformally
invariant version \cite{jordan,flanagan,faraoni,annalen07}.
The point is that these theories can be represented in the two conformally
related frames: the Jordan frame in which the scalar field is
non-minimally coupled to the metric tensor, and in the Einstein
frame in which it is minimally coupled to the metric tensor. Besides, it is
striking that the scalar-tensor theory of gravity is the
low-energy limit of superstring theory \cite{polchinsky,superjim,veneziano,quevedo}.
It has been shown that some physical processes such as the universe inflation
and density perturbations look different in conformally related frames
\cite{veneziano,quevedo}, and so it motivates discussions of these
transformations within the framework of various theories of
gravity.

Many studies have been devoted to the problem of the change of the
geometrical and physical quantities under conformal
transformations (see e.g. \cite{maeda} and references therein). However, the rules of
transformations were not always presented in the most
user-friendly way and also some simple mathematical
properties of these transformations have not been explored up to
very much details. This is why we would like to collect all of these rules
in one paper from the beginning to the end in order to
have a compendium about the conformal transformations. In
particular, we would like to explore the problem of the conformal
transformation properties of the higher-order curvature invariants
as well as conformal transformations as applied to superstring
theory.

Our paper is organized as follows. In Section II we give basic
review of the idea of the conformal transformations of the metric
tensor and discuss the transformation properties of the
geometric quantities such as connection coefficients, Riemann tensor, Ricci tensor
and Ricci scalar. In Section III we discuss
the rules of conformal transformations of the curvature invariants
$R^2$, $R_{ab}R^{ab}$, $R_{abcd}R^{abcd}$
which may emerge as higher-order corrections to standard gravity.
In Setion IV we discuss the conformal transformations of the matter energy-momentum tensor
and emphasize that conformal transformations ``create'' matter.
In Section V we discuss how to
construct conformally invariant theory of gravity. In Section VI we
show the rules of the conformal transformations in the fashion
of duality transformations in superstring theory. In Section VII
we give our conclusions.

\section{Conformal transformations in Einstein gravitation}
\label{conrel}
\setcounter{equation}{0}

Consider a spacetime $({\cal M}, g_{ab})$, where ${\cal M}$ is a smooth
$n-$dimensional manifold and $g_{ab}$ is a Lorentzian metric on $M$.
The following conformal transformation
\bea
\label{conf_trafo}
\tilde{g}_{ab}(x) &=& \Omega^2(x) g_{ab}(x)~,
\eea
where $\Omega$ is a smooth, non-vanishing function of the spacetime point
is a point-dependent rescaling of the metric and is called a conformal factor. It must
lie in the range $0<\Omega<\infty$ ($a,b,k,l = 0,1,2, \ldots D$).
The conformal transformations shrink or stretch the distances between the two points
described by the same coordinate system $x^{a}$ on the manifold
${\cal M}$, but they preserve the angles between vectors
(in particular null vectors which define light cones) which
leads to a conservation of the (global) causal structure of the manifold \cite{hawk_ellis}.
If we take $\Om =$ const. we deal with the so-called scale
transformations \cite{maeda}. In fact, conformal transformations are
localized scale transformations $\Om = \Om(x)$.

On the other hand, the coordinate transformations $x^{a} \to \tilde{x}^{a}$ only
change coordinates and do not change geometry so that
they are entirely different from conformal transformations \cite{hawk_ellis}.
This is crucial since conformal
transformations lead to a different physics \cite{maeda}.
Since this is usually related to a different coupling
of a physical field to gravity, we will be talking about different
frames in which the physics is studied (see also Refs. \cite{flanagan,faraoni} for
a slightly different view).

In $D$ spacetime dimensions the
determinant of the metric $g={\rm det}~[g_{ab}]$ transforms as
\bea
\label{det}
\sqrt{-\tilde{g}} &=& \Omega^D \sqrt{-g}~.
\eea
It is obvious from (\ref{conf_trafo}) that the following relations
for the inverse metrics and the spacetime intervals hold
\bea
\label{conf_trafo_inv}
\tilde{g}^{ab} &=& \Omega^{-2} g^{ab}~, \\
\d \tilde{s}^2 &=& \Omega^2 \d s^2~.
\eea
Finally, the notion of conformal flatness means that
\bea
\label{conf_flat}
\tilde{g}_{ab} \Omega^{-2}(x) &=& \eta_{ab}~,
\eea
where $\eta_{ab}$ is the flat Minkowski metric.

The application of (\ref{conf_trafo}) to the Christoffel connection coefficients
gives (compare \cite{hawk_ellis})
\bea
\label{connections}
\tilde{\Ga}^{c}_{ab}
&=&
\Ga^{c}_{ab} + \frac{1}{\Om}\left( \delta^{c}_{a} \Om_{,b} +
  \delta^{c}_{b} \Om_{,a} - g_{ab}g^{cd}\Om_{,d} \right)~,
  \hspace{0.5cm}\tilde{\Ga}^{b}_{ab}
= \Ga^{b}_{ab} + D \frac{\Om_{,a}}{\Om}
\\
\label{connections1}
\Ga^{c}_{ab}
&=&
\tilde{\Ga}^{c}_{ab}- \frac{1}{\Om} \left( \tilde{\delta}^{c}_{a}
  \Om_{,b} + \tilde{\delta}^{c}_{b} \Om_{,a} -
  \tilde{g}_{ab}\tilde{g}^{cd}\Om_{,d} \right)~,
  \hspace{0.5cm}\Ga^{b}_{ab}
= \tilde{\Ga}^{b}_{ab} - D \frac{\Om_{,a}}{\Om}.
\eea

The Riemann tensors, Ricci tensors, and Ricci scalars in the two related frames $g_{ab}$ and
$\tilde{g}_{ab}$ transform as (we use the sign convention (-+...++), the Riemann
tensor convention $R^a_{~bcd} = \Gamma^a_{bd,c} - \Gamma^a_{bc,d} +
\Gamma^a_{ce} \Gamma^e_{bd} - \Gamma^a_{de} \Gamma^e_{cb}$, and the Ricci tensor is $R_{bd}
= R^{a}_{~bad}$)
\bea
\label{riemanntensor1}
\tilde{R}^a_{~bcd} &=& R^a_{~bcd} + \frac{1}{\Om}\left[\delta^a_d
\Om_{;bc} - \delta^a_c \Om_{;bd} + g_{bc} \Om^{;a}_{~;d} - g_{bd}
\Om^{;a}_{~;c} \right] \\
&+& \frac{2}{\Om^{2}} \left[\delta^a_c
\Om_{,b} \Om_{,d} - \delta^a_d \Om_{,b} \Om_{,c} + g_{bd} \Om^{,a}
\Om_{,c} - g_{bc} \Om^{,a} \Om_{,d} \right]
+ \frac{1}{\Om^{2}} \left[\delta^a_d g_{bc} - \delta^a_c g_{bd}
\right]g_{ef} \Om^{,e} \Om^{,f} ~~, \nonumber
\\
\label{riemanntensor2}
R^a_{~bcd} &=& \tilde{R}^a_{~bcd} - \frac{1}{\Om}\left[\delta^a_d
\Om_{\tilde{;}bc} - \delta^a_c \Om_{\tilde{;}bd} + \tilde{g}_{bc} \Om^{\tilde{;}a}_{~\tilde{;}d}
- \tilde{g}_{bd} \Om^{\tilde{;}a}_{~\tilde{;}c} \right] \\
&+& \frac{1}{\Om^{2}} \left[\delta^a_d \tilde{g}_{bc} - \delta^a_c \tilde{g}_{bd}
\right]\tilde{g}_{ef} \Om^{,e} \Om^{,f} ~~, \nonumber
\eea
\bea
\label{riccitensor1}
\tilde{R}_{ab}
&=&
R_{ab} + \frac{1}{\Om^{2}}\left [
  2(D-2)\Om_{,a}\Om_{,b}-(D-3) \Om_{,c}\Om^{,c}g_{ab}\right ]
-\frac{1}{\Om}\left [ (D-2)\Om_{;ab}+ g_{ab} \Box \Om  \right ]~,
\\
\label{riccitensor2}
R_{ab} &=& \tilde{R}_{ab} - \frac{1}{\Om^2}(D-1)\tilde{g}_{ab} \Om_{,c}\Om^{,c}
+\frac{1}{\Om}\left[(D-2)\Om_{\tilde{;}ab}+ \tilde{g}_{ab} \stackrel{\sim}{\Box} \Om  \right ]~,
\eea
\bea
\label{ricciscalar4}
\tilde{R} &=& \Om^{-2} \left [ R - 2(D-1)\frac{\Box{\Om}}{\Om} -
(D-1)(D-4) g^{ab} \frac{\Om_{,a}\Om_{,b}}{\Om^2}
\right]~,\\
\label{ricciscalar5}
R &=& \Omega^2 \left[ \tilde{R} + 2(D-1)
\frac{\stackrel{\sim}{\Box}\Om}{\Om} - D(D-1) \tilde{g}^{ab}
\frac{\Om_{,a}\Om_{,b}}{\Om^2}\right ]~,
\eea
and the appropriate d'Alambertian operators change under (\ref{conf_trafo}) as
\bea
\label{boxphi1}
\stackrel{\sim}{\Box}\phi &=&\Om^{-2}\left( {\Box}\phi+
  (D-2)g^{ab}\frac{\Om_{,a}}{\Om}\phi_{,b} \right )~,\\
\label{boxphi2}
\Box\phi &=&\Om^{2}\left( \stackrel{\sim}{\Box}\phi-
  (D-2)\tilde{g}^{ab}\frac{\Om_{,a}}{\Om}\phi_{,b} \right )~.
\eea
In these formulas the d'Alembertian $\stackrel{\sim}{\Box}$ taken with respect to the
metric $\tilde{g}_{ab}$ is different
from $\Box$ which is taken with respect to a conformally rescaled metric
$g_{ab}$. Same refers to the covariant derivatives $\tilde{;}$
and $;$ in (\ref{riemanntensor1})-(\ref{riccitensor2}). Note that
some of these quantities are given in Ref. \cite{hawk_ellis} in a
different form. Also, notice that in $D=4$ the rule
(\ref{ricciscalar4}) composes of the two terms only (and it is
often presented in standard textbooks \cite{narlikar} that way),
while the inverse rule (\ref{ricciscalar5}) composes of the three
terms. This reflects the fact that the rules of the simple and
inverse transformations are not symmetric (see Section
\ref{duality} for a discussion of a symmetric, duality-like,
representation).

For the Einstein tensor we have
\bea
\label{Eintensor1}
\tilde{G}_{ab}
&=&
G_{ab} + \frac{D-2}{2\Om^2}\left [
  4\Om_{,a}\Om_{,b}+(D-5) \Om_{,c}\Om^{,c}g_{ab}\right ]
-\frac{D-2}{\Om} \left [ \Om_{;ab} - g_{ab} \Box \Om  \right ]~,
\\
\label{Eintensor2}
G_{ab} &=& \tilde{G}_{ab} + \frac{D-2}{2\Om^2}(D-1)\Om_{,e}\Om^{,e} \tilde{g}_{ab}
+ \frac{D-2}{\Om} \left[\Om_{\tilde{;}ab} - \tilde{g}_{ab} \stackrel{\sim}{\Box}\Om  \right]~,
\eea

An important feature of the conformal transformations is that they
preserve Weyl conformal curvature tensor $(D \geq 3)$
\bea
\label{weyl_def}
C_{abcd} &=& R_{abcd} +
\frac{2}{D-2}\left(g_{a[d}R_{c]b} +
g_{b[c}R_{d]a}\right) + \frac{2}{(D-1)(D-2)} R
g_{a[c}g_{d]b}~,
\eea
which means that we have (note that one index is raised)
\bea
\label{weyl_trafo}
\tilde{C}^{a}_{~bcd} &=& C^{a}_{~bcd}
\eea
under (\ref{conf_trafo}).
Using this property (\ref{weyl_trafo}) and the rules (\ref{conf_trafo})-(\ref{conf_trafo_inv})
one can easily conclude that the Weyl Lagrangian \cite{weyl}
\be
\tilde{L}_w = - \alpha \sqrt{-\tilde{g}} \tilde{C}^{abcd} \tilde{C}_{abcd}
 = - \alpha \sqrt{-g} C^{abcd} C_{abcd} = L_w~
\ee
is an invariant of the conformal transformation
(\ref{conf_trafo}).

\section{Conformal transformations in higher-order gravitation}
\label{conhigher}
\setcounter{equation}{0}

In the physical theories which enter the dense quantum phase of
the evolution of the universe one often applies quantum
corrections to general relativity \cite{birell} which are composed
of the curvature invariants $R^2$, $R_{ab}R^{ab}$, $R_{abcd}R^{abcd}$
and the Gauss-Bonnet invariant $R_{GB}$ \cite{lovelock} as well as their functions such as $f(R)$
\cite{f(R)}, $f(R_{GB})$ \cite{GB} as well as $f(R^2,R_{ab}R^{ab},R_{abcd}R^{abcd})$
\cite{clifton,balc081}. In fact, the additional terms which come from the inclusion of
these curvature invariants play the role of the corrections to the
standard gravity action \cite{weinberg08}. This is why it is useful to
know the rules of the conformal transformations for all these
quantities.

These rules are given as follows
\bea
\label{R2}
\tilde{R}^2 &=& \Om^{-4} \left[ R^2 + 4 (D-1)^2 \Om^{-2} \left( \Box \Om
\right)^2 + (D-1)^2 (D-4)^2 \Om^{-4} g^{ab} \Om_{,a} \Om_{,b}
g^{cd} \Om_{,c} \Om_{,d} \right. \nonumber \\ &-& \left.
4 (D-1) R \Om^{-1} \Box \Om - 2 R
(D-1)(D-4) \Om^{-2} g^{ab} \Om_{,a} \Om_{,b} \right. \nonumber \\
&+& \left. 4(D-1)^2(D-4)
\Om^{-3} \Box \Om g^{ab} \Om_{,a} \Om_{,b} \right]~,
\eea
\bea
\label{RabRab}
\tilde{R}_{ab}\tilde{R}^{ab} &=& \Om^{-4} \left\{ R_{ab}R^{ab} - 2
\Om^{-1} \left[(D-2)R_{ab} \Omega^{;ab} + R \Box \Om \right] \right. \nonumber \\
&+& \left. \Om^{-2} \left[ 4(D-2) R_{ab} \Om^{,a} \Om^{,b} -
2(D-3) R \Om_{,e} \Om^{,e} + (D-2)^2 \Om_{;ab} \Om^{;ab} + (3D -
4) \left( \Box \Om \right)^2 \right] \right. \nonumber \\
&-& \Om ^{-3} \left. \left[ (D-2)^2 \Om_{;ab} \Om^{,a} \Om^{,b} -
(D^2 - 5D + 5) \Box \Om \Om_{,e} \Om^{,e} \right] \right.
\nonumber \\ &+& \left. \Om^{-4} (D-1) (D^2 - 5D + 8) \left(\Om_{,a} \Om^{,a} \right)^2
\right\}~,
\eea
\bea
\label{Riem2}
\tilde{R}_{abcd} \tilde{R}^{abcd} &=& \Om^{-4} \left\{R_{abcd} R^{abcd}
- 8 \Om^{-1} R_{bc} \Omega^{;bc} + 4 \Om^{-2} \left[\left( \Box \Om \right)^2
+ (D-2)\Om_{;bc} \Om^{;bc} - R \Om_{,b} \Om^{,b} \right. \right.
\nonumber \\ &+& \left. \left. 4 R_{bc} \Om^{,b} \Om^{,c}
\right] + 8 \Om^{-3} \left[(D-3) \Box \Om \Om_{,c} \Om^{,c}
- 2(D-2)\Om_{;bc} \Om^{,b} \Om^{,c} \right] \right. \nonumber \\
&+& \left. 2 \Om^{-4} D(D-1) \left(\Om_{,a} \Om^{,a} \right)^2  \right\}~.
\eea

In fact, out of these curvature invariants one forms the well-known Gauss-Bonnet
term which is one of the Euler (or Lovelock) densities \cite{lovelock,GB}. Its
conformal transformation (\ref{conf_trafo}) reads as
\bea
\label{RGBtil}
\tilde{R}_{GB} &\equiv& \tilde{R}_{abcd} \tilde{R}^{abcd} - 4
\tilde{R}_{ab} \tilde{R}^{ab} + \tilde{R}^2
= \Omega^{-4} \left\{ R_{GB} + 4(D-3) \Omega^{-1} \left[ 2
R_{ab} \Omega^{;ab} - R \Box \Om \right] \right. \nonumber \\
&+& \left. 2 (D-3) \Om^{-2} \left[2(D-2) \left( \left( \Box \Om \right)^2
- \Om_{;ab} \Om^{;ab} \right) - 8 R_{ab} \Om^{,a} \Om^{,b} -
(D-6) R \Om_{,a} \Om^{,a} \right] \right. \nonumber \\
&+& \left. 4(D-2)(D-3) \Om^{-3} \left[(D-5) \Box \Om \Om_{,a} \Om^{,a}
+ 4 \Om_{;ab} \Om^{,a} \Om^{,b} \right] \right. \nonumber \\
&+& \left. (D-1)(D-2)(D-3)(D-8) \Om^{-4} \left(\Om_{,a} \Om^{,a} \right)^2
\right\}~.
\eea
The inverse transformation is given by
\bea
\label{RGB}
R_{GB} &\equiv& R_{abcd} R^{abcd} - 4 R_{ab} R^{ab} + R^2 =
\Om^4 \left\{ \tilde{R}_{GB} - 4(D-3) \Om^{-1}
\left[2\tilde{R}_{ab} \Om^{\tilde{;} ab} - \tilde{R} \stackrel{\sim}{\Box} \Om
\right] \right. \nonumber \\
&+& \left. 2(D-2)(D-3) \Om^{-2} \left[ 2\left( \stackrel{\sim}{\Box} \Om
\right)^2 - 2\Om_{\tilde{;}ab} \Om^{\tilde{;}ab} - \tilde{R} \Om_{\tilde{;}a}
\Om^{\tilde{;}a} \right] \right. \\
&-& \left. (D-1)(D-2)(D-3) \Om^{-3} \left[4 \left( \stackrel{\sim}{\Box} \Om
\right) \Om_{\tilde{;}a} \Om^{\tilde{;}a} - D \Om^{-1}
\left( \Om_{\tilde{;}a} \Om^{\tilde{;}a} \right)^2 \right] \right\}~.
\nonumber
\eea

\section{Conformal transformations of the matter energy-momentum tensor }
\label{subtleties}
\setcounter{equation}{0}

So far we have considered only geometrical part. For the
matter part we usually consider the matter action in the form
\bea
\label{matteract}
\tilde{S}_{\rm m} = \int{ \sqrt{-\tilde{g}} d^D x {\tilde{\cal{L}}}_{\rm m} }=
\int{ \sqrt{-g} d^D x {\cal{L}}_{\rm m} }= S_{\rm m}~,
\eea
where the Lagrangians in the conformally related frames transform
as
\be
{\tilde{\cal{L}}}_{\rm m} = \Omega^{-D} {\cal{L}}_{\rm m}
\ee
under the conformal transformation (\ref{conf_trafo}) \cite{superjim}.
Then, the energy-momentum tensor of matter in one conformal frame reads as
\bea
\label{mattertrafotilde}
\tilde{T}_{\rm m}^{ab} &=&
\frac{2}{\sqrt{-\tilde{g}}} \frac{\delta}{\delta \tilde{g}_{ab}}
\left(\sqrt{-\tilde{g}}  {\tilde{\cal{L}}}_{\rm m} \right) =
  \Omega^{-D} \frac{2}{\sqrt{-g}} \frac{\partial g_{cd}}{\partial
  \tilde{g}_{ab}} \frac{\delta}{\delta g_{cd}} \left(\sqrt{-g}
  {\cal{L}}_{\rm m} \right)~,
\eea
which under (\ref{conf_trafo}) transforms as
\bea
\label{tensorlaw1}
\tilde{T}_{\rm m}^{ab} &=& \Omega^{-D-2} T_{\rm m}^{ab}~,\hspace{0.6cm}
\tilde{T}^a_{{\rm m}b} = \Omega^{-D} T^a_{{\rm m}b}, \hspace{0.6cm}
\tilde{T}^{\rm m}_{ab} = \Omega^{-D+2} T^{\rm m}_{ab}~, \hspace{0.6cm}
\tilde{T}_{\rm m} = \Om^{-D} T_{\rm m}~.
\eea
For the matter in the form of the perfect fluid with the
four-velocity $v^{a}$ ($v_{a}v^{b}=-1$), the energy density $\vr$
and the pressure ${\rm p}$
\bea
\label{enmom}
T_{\rm m}^{ab} &=& (\vr+{\rm p})v^{a}v^{b}+{\rm p}g^{ab}~,
\eea
the conformal transformation gives
\bea
\label{enmom_trafo0}
\tilde{T}_{\rm m}^{ab}
&=&
(\tilde{\vr}+\tilde{{\rm p}})\tilde{v}^{a}\tilde{v}^{b} +
\tilde{{\rm p}}\tilde{g}^{ab}~,
\eea
where
\bea
\label{mattertrafo}
T_{\rm m}^{ab} &=& \frac{2}{\sqrt{-g}}\frac{\delta}{\delta  g_{ab}}\left(
  \sqrt{-g}{\cal{L}}_{\rm m}\right )~,
\eea
and
\bea
\tilde{v}^{a} &=& \frac{\d x^{a}}{\d \tilde{s}} =
\frac{1}{\Om}\frac{\d x^{a}}{\d s}=\Om^{-1}v^{a}~.
\eea
Therefore, the relation between the pressure and the energy density in the conformally
related frames reads as
\bea
\tilde{\vr} &=& \Om^{-D} \vr~,\\
\tilde{{\rm p}} &=& \Om^{-D} {\rm p}~.
\eea
It is easy to note that the imposition of the conservation law in the first frame
\bea
\label{laws0}
T^{ab}_{~{\rm m};b}=0~,
\eea
gives in the conformally related frame
\bea
\label{tildelaws0}
 \tilde{T}^{ab}_{~{\rm m}\tilde{;}b} &=& -\frac{\Om^{,a}}{\Om}\tilde{T}_{\rm m}~.
\eea
From (\ref{tildelaws0}) it appears obvious that the conformally
transformed energy-momentum tensor is conserved only, if the trace of
it vanishes ($\tilde{T}_{\rm m}=0$) \cite{jordan,weinberg,maeda,annalen07}. For example,
in the case of barotropic fluid with
\bea
\label{barotropic}
{\rm p}&=&(\ga-1)\vr~ \hspace{0.5cm} \gamma = {\rm const.},
\eea
it is conserved only for the radiation-type fluid
${\rm p}=[1/(D-1)]\vr$~.

Similar considerations are also true if we first impose the
conservation law in the second frame
\bea
\label{laws10}
\tilde{T}^{ab}_{~{\rm m}\tilde{;}b}=0~,
\eea
which gives in the conformally related frame (no tildes)
\bea
\label{tildelaws10}
T^{ab}_{~{\rm m};b} &=& \frac{\Om^{,a}}{\Om}T_{\rm m}~.
\eea
Finally, it follows from (\ref{tensorlaw1}) that that vanishing of
the trace of the energy-momentum tensor in one frame necessarily
requires its vanishing in the second frame, i.e., if $T_{\rm m}=0$ in one frame, then
$\tilde{T}_{\rm m}=0$ in the second frame and vice versa. This means only the traceless type
of matter fulfills the requirement of energy conservation.

We now use the formulas (\ref{Eintensor1})-(\ref{Eintensor2}) to
discuss the formulation of the Einstein field equations for the
case of the conformal transformations. Let us assume then the
validity of the Einstein field equations in one of the conformal
frames (no tildes) as follows
\bea
\label{Gab}
G_{ab} &=& \kappa^2 T^{\rm m}_{ab}, \hspace{1.cm} G^{ab} = \kappa^2
T_{\rm m}^{ab}~,
\eea
so that the imposition of the Bianchi identity gives
\bea
\label{GabBia}
G^{ab}_{~~;b} &=& 0 \Rightarrow T^{ab}_{{\rm m}~;b} = 0~,
\eea
However, due to (\ref{Eintensor2}) in a conformally related frame
(with tildes) one has
\bea
\label{EinNT}
G_{ab} &=& \tilde{G}_{ab} + \tilde{T}^{\Omega}_{ab}, \hspace{1.cm}
G^{ab} = \Omega^4 \left( \tilde{G}^{ab} + \tilde{T}_{\Omega}^{ab}
\right)~,
\eea
where
\bea
\label{tT_om}
\tilde{T}^{\Omega}_{ab} &=& -\frac{D-2}{2\Om^2}(D-1)\Om_{,e}\Om^{,e} \tilde{g}_{ab}
+ \frac{D-2}{\Om} \left[\Om_{\tilde{;}ab} - \tilde{g}_{ab} \stackrel{\sim}{\Box}\Om
\right]~.
\eea
Using these one can write down the Einstein equations (\ref{Gab})
as
\bea
\label{tEA1}
\tilde{G}_{ab} + \tilde{T}_{ab}^{\Omega} = \kappa^2 \Om^{D-2}
\tilde{T}^m_{ab}~,
\eea
or, alternatively as
\bea
\label{tEA2}
\tilde{G}^{ab} = \kappa^2 \Om^{D+2} \tilde{T}^{ab}_m -
\tilde{T}^{ab}_{\Om}~.
\eea
Left-hand side of (\ref{tEA2}) is geometrical, so that the
imposition of the Bianchi identity gives
\bea
\label{tEA3}
\tilde{G}^{ab}_{~~\tilde{;}b} = 0 = \kappa^2 (\Om^{D+2}
\tilde{T}^{ab}_m )_{\tilde{;}\hspace{1.pt}b} - \tilde{T}_{\Om \hspace{1.pt}
\tilde{;}\hspace{1.pt} b}^{ab}~.
\eea
Using (\ref{tensorlaw1}), (\ref{laws0}) and (\ref{tildelaws0}) after some manipulations one
gets
\bea
\label{integ1}
\kappa^2 \left[(D+2) \frac{\Om_{,b}}{\Om} T_m^{ab} -
\Om^{,a}\Om T_m \right] = \tilde{T}^{ab}_{\Om \hspace{1.pt} \tilde{;}
\hspace{1.pt} b}~.
\eea
The right-hand side of (\ref{integ1}) can be obtained after
differentiating of (\ref{tT_om}). The conclusion is that we need
to take into account an extra matter term in (\ref{tEA2}) which
comes from conformal factor so that one may read (\ref{tEA2}) in
the form
\bea
\tilde{G}^{ab} = \kappa^2 \tilde{T}^{ab}_{total}~,
\eea
and
\bea
\tilde{T}^{ab}_{total} = \hat{T}^{ab}_{m} - \frac{1}{\kappa^2}
\tilde{T}^{ab}_{\Om}~,
\eea
where
\bea
\hat{T}^{ab}_{m} = \Om^{D+2} \tilde{T}^{ab}_m~.
\eea
Now, let us make the transformation in an inverse way. Assume
\bea
\label{tGab}
\tilde{G}_{ab} &=& \kappa^2 \tilde{T}^{\rm m}_{ab}, \hspace{1.cm} \tilde{G}^{ab} = \kappa^2
\tilde{T}_{\rm m}^{ab}~,
\eea
so that the imposition of the Bianchi identity gives
\bea
\label{tGabBia}
\tilde{G}^{ab}_{~~\tilde{;}~b} = 0 \Rightarrow \tilde{T}^{ab}_{{\rm m}~\tilde{;}~b} = 0~,
\eea
In a conformally related frame
(without tildes) one has
\bea
\label{tEinNT}
\tilde{G}_{ab} &=& G_{ab} + T^{\Omega}_{ab}, \hspace{1.cm}
\tilde{G}^{ab} = \Omega^{-4} \left( \tilde{G}^{ab} + \tilde{T}_{\Omega}^{ab}
\right)~,
\eea
where
\bea
\label{T_om}
T^{\Om}_{ab} = - \frac{D-2}{2\Om^2}
\left[4\Om_{,a}\Om_{,b}+(D-5) \Om_{,c}\Om^{,c}g_{ab}\right ]
-\frac{D-2}{\Om} \left [ \Om_{;ab} - g_{ab} \Box \Om  \right ]~.
\eea
Now, the field equations (\ref{tGab}) give
\bea
\label{Gabfin}
G_{ab} + T^{\Om}_{ab} = \kappa^2 \Om^{-D-2} T^{\rm m}_{ab}~,
\eea
or, alternatively
\bea
G^{ab} = \kappa^2 \Om^{-D+2} T_{\rm m}^{ab} - T^{ab}_{~\Om}~,
\eea
and the Bianchi identity gives
\bea
G^{ab}_{~~;b} &=& 0 = \kappa^2 \left(\Omega^{-D+2} T^{ab}_{~{\rm
m}} \right)_{;b} - T^{ab}_{~\Omega;b}~.
\eea
In order to be consistent with the previous derivation we assume
(\ref{laws0}) which finally gives
\bea
\kappa^2 (-D+2) \Om^{-D+2} \frac{\Om_{,b}}{\Om} T^{ab}_{~\rm m} =
T^{ab}_{~\Om;b}~.
\eea

As an amazing example of the subtlety of the conformal
transformation applied to matter energy-momentum tensor let us
consider the ``creation'' of the Friedmann universes out of the
flat Minkowski spacetime. In order to do that
we start with the flat Minkowski spacetime with a flat Minkowski metric
$g_{ab} = \eta_{ab}$ and $\tilde{g}_{ab} = \Omega^2 \eta_{ab}$. Assuming no matter
energy-momentum tensor we have from (\ref{tensorlaw1}) that $T_{\rm m}^{ab} = 0$
which implies $\tilde{T}_{\rm m}^{ab} = 0$. Then, from
(\ref{tEA2}) we have that despite $G_{ab} = 0$ one has the matter
``created'', i.e.,
\bea
\tilde{G}^{ab} = - \tilde{T}^{ab}_{\Om} \neq 0~.
\eea
For the flat Friedmann universe being ``created'' out of the flat
Minkowski universe one has ($a,b = 0,1,2,3$)
\bea
ds^2 = a^2(\eta)(-d\eta^2 + dx^2 + dy^2 + dz^2) =
\Omega^2(\eta) \eta_{ab} dx^a dx^b~,
\eea
and the conformal factor $\Omega = a(\eta)$ is equal to the scale factor while
$\eta$ is the conformal time. In this simple case one has
\bea
\tilde{T}^{0}_{0~\Om} = - 3 \frac{\dot{a}}{a^4}, \hspace{0.3cm}
\tilde{T}^{\mu}_{\mu~\Om} =  \frac{1}{a^4} \left[2 a \ddot{a} -
\dot{a}^2 \right]~,
\eea
where $\mu = 1,2,3$ and the dot represents the derivative with respect to conformal
time $\eta$. The other components of the tensor $\tilde{T}_a^b$
are zero.

One can give the following physical interpretation of the effect
under study. The energy-momentum tensor in a Friedmann universe is
``created'' out of the flat Minkowski
space due to the work done by the conformal transformation to bend
flat space and become curved.

\section{Conformally invariant gravitation}
\label{confinvsect}
\setcounter{equation}{0}

Let us start with the vacuum Einstein-Hilbert action of general relativity
in $D$ spacetime dimensions which read as (in $D=4$ dimensions $\xi = 1/6$
and $\kappa^2 = 8\pi G \equiv 6$)
\bea
\label{ehconf1D}
{\S}_{EH} &=& \frac{\xi}{2}\int~\d^Dx\stg \tilde{R}~,
\eea
where
\bea
\xi &=& \frac{1}{4}\frac{D-2}{D-1}~.
\eea
This is the so-called Einstein frame action.
The application of the formula (\ref{ricciscalar4}) to
(\ref{ehconf1D}) gives a new action
\bea
\label{ehconf11D}
{\S}_{EH} &=& \frac{1}{2}\int~\d^Dx\sg \xi \Om^{D-2} \left[ R -
2(D-1) \frac{\Box{\Om}}{\Om} -
  \frac{(D-1)(D-4)}{\Om^2} g^{\mu\nu} \Om_{,\mu}\Om_{,\nu}\right
  ]~,
\eea
which is not conformally invariant, apart from
the case of the global transformations of the trivial type
$\tilde{g}_{\mu\nu} = {\rm const}. \times g_{\mu\nu}$.
However, if we start with the action (Jordan frame action)
\bea\label{ehconf2D}
{\tilde{\S}}_{C} &=&
\frac{1}{2}\int~\d^Dx\stg \xi \tilde{R}\tilde{\Phi}^2~,
\eea
then the result of the conformal transformation will be as follows
\bea
\label{ehconf21D}
{\tilde{\S}}_{C} &=& \frac{1}{2}\int~\d^Dx\sg \Om^{D-2}\tilde{\Phi}^2 \left[ \xi R -
\frac{(D-2)}{2} \frac{\Box{\Om}}{\Om} -
  \frac{(D-2)(D-4)}{4\Om^2} g^{\mu\nu} \Om_{,\mu}\Om_{,\nu}\right
  ]~,
\eea
so that we can redefine the scalar field as
\bea
\label{PhitildetoPhiD}
\tilde{\Phi} &=& \Om^{\frac{2-D}{2}} \Phi~
\eea
to get
\bea
\label{ehconf212D}
{\tilde{\S}}_{C} &=& \frac{1}{2}\int~\d^Dx\sg \Phi^2 \left[ \xi R -
\frac{(D-2)}{2} \frac{\Box{\Om}}{\Om} -
  \frac{1}{4}(D-2)(D-4) g^{\mu\nu} \frac{\Om_{,\mu}\Om_{,\nu}}{\Om^2} \right
  ]~.
\eea
Now we add
\bea
\label{field_act1D}
{\S}_{\tilde{\Phi}} &=&-\frac{1}{2}\int~\d^Dx\stg
\tilde{\Phi}\stackrel{\sim}{\Box}\tilde{\Phi}~.
\eea
First we use (\ref{boxphi1}) with $\phi = \tilde{\Phi}$ to get
\bea
\label{boxPhi1tilde}
\stackrel{\sim}{\Box} {\tilde{\Phi}} &=& \Om^{-2} \left [
  {\Box}\tilde{\Phi} +
  (D-2)g^{\mu\nu}\frac{\Om_{,\mu}}{\Om}\tilde{\Phi}_{,\nu}
  \right]~.
\eea
Then, from (\ref{PhitildetoPhiD}) we have
\bea
\label{boxtildePhi}
\Box{\tilde{\Phi}} &=& \frac{D(D-2)}{4} \Om^{-\frac{D+2}{2}}
g^{\mu\nu} \Om_{,\mu} \Om_{,\nu} \Phi + \frac{2-D}{2}
\Om^{-\frac{D}{2}} \Phi \Box \Om \nonumber \\
&-& (D-2) \Om^{-\frac{D}{2}}
\Om_{,\mu} \Phi_{,\nu} g^{\mu\nu} + \Om^{\frac{2-D}{2}} \Box \Phi~.
\eea
The derivatives of $\tilde{\Phi}$ give
\bea
\label{tildePhicov}
\tilde{\Phi}_{,\nu} &=& \frac{2-D}{2} \Om^{-\frac{D}{2}}
\Om_{,\nu} \Phi + \Om^{\frac{2-D}{2}} \Phi_{,\nu}~,\\
\tilde{\Phi}^{,\mu} &=& \tilde{g}^{\mu\nu} \tilde{\Phi}_{,\nu} =
\frac{1}{\Om^2} \left[ \frac{2-D}{2} \Om^{-\frac{D}{2}}
\Om^{,\mu} \Phi + \Om^{\frac{2-D}{2}} \Phi^{,\mu} \right]~.
\eea
On the other hand, from (\ref{boxPhi1tilde}) and (\ref{boxtildePhi}) we get
\bea
\label{box2tilde}
\stackrel{\sim}{\Box} {\tilde{\Phi}} &=& \Om^{-2} \left [
\frac{2-D}{2} \Om^{-\frac{D}{2}} \Phi \Box \Om + \Om^{\frac{2-D}{2}}
\Box \Phi - \frac{(D-2)(D-4)}{4} \Om^{-\frac{D+2}{2}}
g^{\mu\nu}\Om_{,\mu}\Om_{,\nu} \Phi \right]~,
\eea
which after the substitution into (\ref{field_act1D}) gives
\bea
\label{ehconf3D}
{\S}_{\tilde{\Phi}} &=& -\frac{1}{2} \int d^D x \sg \left[ \frac{2-D}{2} \Phi^2
\frac{\Box{\Om}}{\Om} + \Phi \Box{\Phi} - \frac{1}{4}(D-2)(D-4)
g^{\mu\nu} \frac{\Omega_{,\mu}\Omega_{,\nu}}{\Om^2} \Phi^2
\right]~.
\eea
Now we notice that the total action in an original frame
\bea
\label{tildeconfinvD}
\tilde{\S} ={\S}_{\tilde{\Phi}}+{\tilde{\S}}_{C}
&=&
 \frac{1}{2}\int~\d^Dx\stg\tilde{\Phi}\left (
  \frac{1}{4}\frac{D-2}{D-1} \tilde{R}\tilde{\Phi} -\stackrel{\sim}{\Box}\tilde{\Phi}\right )~,
\eea
is, in fact, conformally invariant, since
\bea
\label{confinvD}
\S = {\S}_{\Phi} + S_C &=&
 \frac{1}{2}\int~\d^Dx\sg{\Phi}\left (\frac{1}{4}\frac{D-2}{D-1}{R}{\Phi}-{\Box}\Phi \right
 )~,
\eea
with $S_{\Phi}$ and $S_C$ defined for the quantities without
tildes. The conformally invariant actions (\ref{tildeconfinvD}) and
(\ref{confinvD}) are the basis to derive the equations of motion via the variational
principle. The scalar field equations of motion are
\bea
\label{eom1D}
\left
  (\stackrel{\sim}{\Box}-\frac{1}{4}\frac{D-2}{D-1} \tilde{R} \right ) \tilde{\Phi} =
   \Om^{-\frac{D+2}{2}} \left(\Box - \frac{1}{4}\frac{D-2}{D-1} R\right)
  \Phi &=& 0~,
\eea
and they are also conformally invariant having the structure of the Klein-Gordon
equation with the mass term replaced by the curvature term ~\cite{chern}.

The conformally invariant field equations in D dimensions are
\bea
\label{eom2D}
\left ( \tilde{R}_{\mu\nu}-\frac{1}{2}\tilde{g}_{\mu\nu}\tilde{R}\right)
\frac{1}{4}\frac{D-2}{D-1} \tilde{\Phi}^2 + \tilde{\Phi}_{,\mu}\tilde{\Phi}_{,\nu} -
\frac{1}{2}\tilde{g}_{\mu\nu}
\tilde{\Phi}_{,\al}\tilde{\Phi}^{,\al} + \frac{1}{4} \frac{D-2}{D-1} \left[ \tilde{g}_{\mu\nu}
\tilde{\Phi} \stackrel{\sim}{\Box} (\tilde{\Phi}^2) - \tilde{\Phi}
(\tilde{\Phi}^2)_{\tilde{;}\mu\nu} \right] &=& 0~.\nonumber \\
\eea
Since
\bea
\label{box2}
\stackrel{\sim}{\Box} \left( \tilde{\Phi}^2 \right) &=&
2 \tilde{\Phi}_{,\al} \tilde{\Phi}^{,\al} + 2 \tilde{\Phi}
\stackrel{\sim}{\Box} \tilde{\Phi}~, \\
\label{box3}
\left( \tilde{\Phi}^2 \right)_{\tilde{;}\mu\nu} &=&
2 \tilde{\Phi}_{,\mu} \tilde{\Phi}_{,\nu} + 2 \tilde{\Phi}
\tilde{\Phi}_{\tilde{;}\mu\nu}~,
\eea
then by using (\ref{box2})-(\ref{box3}) the equations (\ref{eom2D}) can be cast to
\bea
\label{eom3D}
\left( \tilde{R}_{\mu\nu}- \frac{1}{2}\tilde{g}_{\mu\nu}\tilde{R}
\right) \frac{1}{4} \frac{D-2}{D-1} \tilde{\Phi}^2 &+& \frac{1}{2(D-1)}
\left[ D \tilde{\Phi}_{,\mu}\tilde{\Phi}_{,\nu} -
\tilde{g}_{\mu\nu}
\tilde{\Phi}_{,\al}\tilde{\Phi}^{,\al} \right] \nonumber \\
&+& \frac{1}{2} \frac{D-2}{D-1} \left[ \tilde{g}_{\mu\nu}
\tilde{\Phi} \stackrel{\sim}{\Box} \tilde{\Phi} -
\tilde{\Phi} \tilde{\Phi}_{\tilde{;}\mu\nu} \right] =
0~.
\eea
Notice that the scalar field equations of motion
(\ref{eom1D}) can be obtained by the contraction of (\ref{eom2D}) or (\ref{eom3D}).
In order to prove the conformal invariance of the field equations (\ref{eom3D})
it is necessary to know the rule of the conformal
transformations for the twice covariant derivative of a scalar field
which reads as
\bea
\label{covdertildeD}
\tilde{\Phi}_{\tilde{;}\mu\nu} &=& \tilde{\Phi}_{,\mu\nu} - \tilde{\Ga}^{\rho}_{\mu\nu}
\tilde{\Phi}_{,\rho} = - \frac{1}{2} (D-2) \Om^{-\frac{D}{2}} \Phi \Om_{;\mu\nu} +
\Om^{\frac{2-D}{2}} \Phi_{;\mu\nu} + \frac{1}{4} (D-2)(D+4) \Om^{-\frac{D+2}{2}}
\Phi \Om_{,\mu} \Om_{,\nu} \nonumber \\
&-& \frac{D}{2} \Om^{-\frac{D}{2}}
\left(\Phi_{,\mu}\Om_{,\nu} + \Om_{,\mu} \Phi_{,\nu} \right) -
\frac{1}{2} (D-2) \Om^{-\frac{D+2}{2}} \Phi g_{\mu\nu} \Om_{,\rho} \Om^{,\rho} +
\Om^{-\frac{D}{2}} g_{\mu\nu} \Phi_{,\rho} \Om^{,\rho} ~.
\eea
Inserting (\ref{riccitensor1}), (\ref{ricciscalar4}), (\ref{tildePhicov}),
(\ref{box2tilde}) and (\ref{covdertildeD}) into (\ref{eom3D}) one is able to
prove the conformal invariance of the gravitational equations of
motion which now have the same form in a conformally related frame, i.e.,
\bea
\label{eom4D}
\left( R_{\mu\nu}- \frac{1}{2}g_{\mu\nu}R
\right) \frac{1}{4} \frac{D-2}{D-1} \Phi^2 &+& \frac{1}{2(D-1)}
\left[ D \Phi_{,\mu}\Phi_{,\nu} -
g_{\mu\nu}
\Phi_{,\al} \Phi^{,\al} \right] \nonumber \\
&+& \frac{1}{2} \frac{D-2}{D-1} \left[ g_{\mu\nu}
\Phi \Box \Phi -
\Phi \Phi_{;\mu\nu} \right] = 0~.
\eea
The field equations (\ref{eom3D}) and (\ref{eom4D}) generalize those of
Hoyle-Narlikar theory onto an arbitrary number of spacetime dimensions.
For $D=4$ these are exactly the same field equations as in the Hoyle-Narlikar theory
\cite{HN,narlikar}. Note that the scalar field equations of motion (\ref{eom1D})
can be obtained by the appropriate contraction of equations (\ref{eom3D}) and (\ref{eom4D})
so that they are not independent and do not supply any additional
information \cite{canuto}.

The actions (\ref{tildeconfinvD}) and (\ref{confinvD}) are
usually represented in a different form by the application of the expression
for a covariant d'Alambertian for a scalar field in general relativity
\bea
\label{boxpartial}
\stackrel{\sim}{\Box}\tilde{\Phi} &=& \frac{1}{\sqrt{-\tilde{g}}}
\stackrel{\sim}{\partial}_{a} \left(\sqrt{-\tilde{g}}
\stackrel{\sim}{\partial}^{a}{\tilde{\Phi}}
\right)~,
\eea
which after integrating out the boundary term, gives \cite{maeda}
\bea
\label{boundarytilde}
\tilde{S} &=& \frac{1}{2} \int d^4x \sqrt{-\tilde{g}} \left[
\xi \tilde{R} \tilde{\Phi}^2 +
\stackrel{\sim}{\partial}_{a}\tilde{\Phi}\stackrel{\sim}{\partial}^{a}\tilde{\Phi}
\right]~,
\eea
and the second term is just a kinetic
term for the scalar field (cf. \cite{birell,hawk_ellis}). The
equations (\ref{boundarytilde}) are also conformally invariant since
the application of the formulas (\ref{det}), (\ref{ricciscalar4})
and (\ref{PhitildetoPhiD}) together with the appropriate
integration of the boundary term gives the same form of the equations
\bea
\label{boundary}
S &=& \frac{1}{2} \int d^4x \sqrt{-g} \left[
\xi R \Phi^2 +
{\partial}_{a}\Phi {\partial}^{a}{\Phi}
\right]~.
\eea
Because of the type of non-minimal coupling of gravity to a scalar
field $\tilde{\Phi}$ or $\Phi$ in (\ref{boundarytilde}) or
(\ref{boundary}) appropriately and the relation to Brans-Dicke theory
we say that these equations are presented in the Jordan frame \cite{jordan,maeda}.
It is interesting that the action (\ref{boundary}) for $D=4$ is just the
Brans-Dicke action \cite{bd} with the Brans-Dicke parameter $\omega=-3/2$
\cite{annalen07}.

Note that in $D=4$ dimensions the action (\ref{tildeconfinvD}) reads as
\bea
\label{tildeconfinv}
\tilde{\S} &=&
 \frac{1}{2}\int~\d^4x\stg\tilde{\Phi}\left (
  \frac{1}{6}\tilde{R}\tilde{\Phi} -\stackrel{\sim}{\Box}\tilde{\Phi}\right )~,
\eea
together with (\ref{PhitildetoPhiD}) as
\bea
\label{PhitildetoPhi}
\tilde{\Phi} &=& \Om^{-1} \Phi
\eea
which, in fact, is conformally invariant, since the conformally
transformed action has the same form, i.e.,
\bea
\label{confinv}
\S &=&
 \frac{1}{2}\int~\d^4x\sg{\Phi}\left (\frac{1}{6}{R}{\Phi}-{\Box}\Phi \right
 )~.
\eea
Now, one can see that the original form of the Einstein-Hilbert
action (\ref{ehconf1D}) can be recovered from (\ref{tildeconfinv}) (or, alternatively
(\ref{confinv})) provided we assume that
\bea
\label{kappaphi}
\kappa^2 &=& \frac{6}{\tilde{\Phi}^2} = {\rm const.}
\eea

Let us now notice that we can formally attach an energy-momentum
tensor of the scalar field in both frames writing (one gets these
expressions by putting $\Omega = \tilde{\Phi}$ in (\ref{tT_om}) and
$\Omega = \Phi$ in (\ref{T_om}))
\bea
\label{Tmntilde}
^{\tilde{\Phi}}\tilde{T}_{ab}  &\equiv&
\frac{1}{6}\left[ \tilde{g}_{ab} \tilde{\Phi}_{,c}\tilde{\Phi}^{,c} -
4 \tilde{\Phi}_{,a}\tilde{\Phi}_{,b} \right]
+
\frac{1}{3} \tilde{\Phi} \left[ \tilde{\Phi}_{\tilde{;}ab} - \tilde{g}_{ab}
\stackrel{\sim}{\Box} \tilde{\Phi} \right]~,
\eea
and
\bea
\label{Tmn}
^{\Phi}T_{ab} &\equiv&
\frac{1}{6}\left[ g_{ab}
\Phi_{,c} \Phi^{,c} - 4 \Phi_{,a}\Phi_{,b} \right]
+
\frac{1}{3}\Phi \left[ \Phi_{;ab} - g_{ab} \Box \Phi
\right]~,
\eea
which allow to brief the equations (\ref{eom3D}) and (\ref{eom4D})
to the familiar Einstein form
\bea
\label{Biatilde}
\tilde{G}_{ab} &=& \tilde{R}_{ab}- \frac{1}{2}\tilde{g}_{ab}\tilde{R} =
\frac{6}{\tilde{\Phi}^2} \left(^{\tilde{\Phi}}\tilde{T}_{ab}\right) \equiv
 ^{con\tilde{\Phi}} \tilde{T}_{ab} = \tilde{T}^{\Omega}_{ab}~,
\eea
and
\bea
\label{Bia}
G_{ab} = R_{ab}- \frac{1}{2}g_{ab}R &=& \frac{6}{\Phi^2} \left(^{\Phi}T_{ab}\right)
\equiv  ^{con\Phi} T_{ab}= T^{\Omega}_{ab}~,
\eea
and we have added the abbreviation ``con'' to mark the fact that
these tensors are conserved due to the Bianchi identity (compare with (\ref{tT_om}) and
(\ref{T_om})). Notice that the contraction of (\ref{Tmn}) gives
\bea
\left(^{\Phi} T \right)&=& - \Phi \Box{\Phi}~,
\eea
which in view of (\ref{eom1D}) gives
the condition
\bea
\label{RT}
R &=& -\frac{6}{\Phi^2} \left(^{\Phi} T \right) = 6 \frac{\Box \Phi}{\Phi}~,
\eea
which is just a contraction of (\ref{Bia}).

If so, by the application of Bianchi identity to the left-hand
sides of (\ref{Biatilde}) and (\ref{Bia}) we formally get the
conservation laws for these conserved energy-momentum tensors, i.e.,
\bea
^{con \tilde{\Phi}}\tilde{T}^{a}_{~b \tilde{;}a} &=&0~,
\eea
and
\bea
^{con \Phi}T^{a}_{~b ; a} &=&0~.
\eea
Writing down the conservation law as if it referred to tensors
$^{\tilde{\Phi}}\tilde{T}^{ab}$ and $^{\Phi}T^{ab}$ gives
\bea
^{\tilde{\Phi}}\tilde{T}^{a}_{~b \tilde{;} a} &=& 2
\frac{\tilde{\Phi}_{,a}}{\tilde{\Phi}} \hspace{3pt} ^{\tilde{\Phi}} \tilde{T}^{a}_{~b} =
\frac{1}{3} \tilde{\Phi} \tilde{\Phi}_{,a} \left(\tilde{R}^{a}_{b} -
\frac{1}{2} \delta^{a}_{b} \tilde{R} \right)~,
\eea
and
\bea
^{\Phi}T^{a}_{~b ;a} &=& 2 \frac{\Phi_{,a}}{\Phi} \hspace{0pt} ^{\Phi}T^{a}_{b}
= \frac{1}{3} \Phi \Phi_{,a} \left(R^{a}_{b} - \frac{1}{2} \delta^{a}_{b} R
\right)~,
\eea
and they look like the matter was created.

Another point is that the equations (\ref{eom3D}) or (\ref{eom4D}) apparently could
give directly the vacuum Einstein field
equations for $\tilde{\Phi}= \sqrt{6}/\kappa = \sqrt{6/8\pi G} =$ const. (cf. Eq.
(\ref{kappaphi})). The same is obviously true for the field
equations (\ref{eom4D}) with the same value of $\Phi = \sqrt{6}/\kappa =
\sqrt{6/8\pi G}$ = const. However, this limit is restricted to the
case of vanishing Ricci curvature $R=0$ or $\tilde{R}=0$ (so only flat Minkowski space limit is
allowed) which can
be seen from the scalar field equations of motion (\ref{eom1D}).

The admission of the matter part (\ref{matteract})
into the action (\ref{confinv}) allows to generalize the field equations (\ref{eom4D}) to
\bea
\label{eom4T}
\left( R_{ab}- \frac{1}{2}g_{ab}R
\right) \frac{1}{6} \Phi^2 + \frac{1}{6}
\left[ 4 \Phi_{,a}\Phi_{,b} -
g_{ab}
\Phi_{,c} \Phi^{,c} \right] + \frac{1}{3} \left[ g_{ab}
\Phi \Box \Phi -
\Phi \Phi_{;ab} \right] &=&   T^{\rm m}_{ab}~,
\eea
or
\bea
\label{eom4T1}
G_{ab} &=& R_{ab}- \frac{1}{2}g_{ab}R = \frac{6}{\Phi^2}
\left[^{\Phi} T_{ab} + T^{\rm m}_{ab} \right]~,
\eea
and this last formula (\ref{eom4T1}) is equivalent to
(\ref{Gabfin}).
These equations (\ref{eom4T1}), after contraction, give modified field equations
(\ref{eom1D})
\bea
\label{eom_1T}
\left(\Box - \frac{1}{6} R \right) \Phi &=&
\frac{T_{\rm m}}{\Phi}~,
\eea
where we have used a generalized relation (\ref{RT})
\bea
\label{RT1}
R &=& -\frac{6}{\Phi^2} \left(^{\Phi} T + T_{\rm m} \right)~.
\eea
However, the variation of the action with matter term which does not depend on the
scalar field $\Phi$, i.e.,
\bea
{\cal L}_{\rm m} &\neq& {\cal L}_{\rm m} (\Phi)
\eea
leaves the scalar field equation of motion (\ref{eom1D}) {\it intact}.
In view of (\ref{eom_1T}) this necessarily requires that the trace
of the energy-momentum tensor of matter must vanish, i.e.,
\bea
\label{T=0}
T_{\rm m} &=& 0 .
\eea
It means that only the traceless type of matter can be made consistent with
the matter energy-momentum tensor independent of the scalar field
$\Phi$. This fact is often expressed as a statement that "photons
weigh, but the Sun does not" \cite{barber}.

In a conformally related frame we apply the matter term
(\ref{matteract}) and (\ref{mattertrafotilde}) to get
\bea
\label{eom4Ttil}
\left( \tilde{R}_{ab}- \frac{1}{2}\tilde{g}_{ab}\tilde{R}
\right) \frac{1}{6} \tilde{\Phi}^2 + \frac{1}{6}
\left[ 4 \tilde{\Phi}_{,a}\tilde{\Phi}_{,b} -
\tilde{g}_{ab}
\tilde{\Phi}_{,c} \tilde{\Phi}^{,c} \right] + \frac{1}{3} \left[ \tilde{g}_{ab}
\tilde{\Phi} \Box \tilde{\Phi} -
\tilde{\Phi} \tilde{\Phi}_{;ab} \right] &=&  \tilde{T}^m_{ab}~,
\eea
or (cf. (\ref{tEA1}))
\bea
\label{eom4Ttil1}
\tilde{G}_{ab} = \tilde{R}_{ab}- \frac{1}{2}\tilde{g}_{ab} \tilde{R}
&=&  \frac{6}{\tilde{\Phi^2}} \left[^{\Phi} \tilde{T}_{ab} +
\tilde{T}^m_{ab}\right]~,
\eea
which after contraction give a modified equation (\ref{eom1D})
\bea
\label{eom_1Ttil}
\left(\stackrel{\sim}{\Box} - \frac{1}{6} \tilde{R} \right) \tilde{\Phi} &=&
\frac{\tilde{T}_m}{\tilde{\Phi}}~.
\eea

From (\ref{tildelaws0}) it appears transparent that the conformally
transformed energy-momentum tensor is conserved only if the trace of
it vanishes ($\tilde{T}=0$) \cite{jordan,weinberg}. For example,
in the case of barotropic fluid (\ref{barotropic})
it vanishes only for radiation ${\rm p}=(1/3)\vr$~. This means that only the photons may
obey the equivalence principle and this is not the case for other types
of matter since with non-vanishing trace in (\ref{tildelaws0}) we
deal with creation of matter process (compare with Self Creation Cosmology of Ref.
\cite{barber} which has the same field equations (\ref{eom4Ttil})
and (\ref{tildelaws0}), but the equation (\ref{eom_1Ttil}) is the
same only for a vanishing curvature scalar $\tilde{R}$).

The tracelessness of the energy-momentum tensor as a result of the imposition of the conservation
law saves the equivalence principle - conformally invariant matter follows geodesic
trajectories. An interesting solution is to allow both traceless
and traceful matter in which radiation fulfills the equivalence
principle and follow geodesics while ordinary matter does not
\cite{barber}. We will come to this problem later. Of course, in
that case ordinary matter is "created" during the evolution and
the conservation law is not fulfilled. This will also happen in
our case which is alike in Hoyle-Narlikar theory \cite{HN}.

Let us notice that the application of the Bianchi identity to
(\ref{eom4T1}) gives
\bea
\label{conserv1}
^{\Phi}T^{a}_{~b ; a} &=& 2 \frac{\Phi_{,a}}{\Phi} \left( ^{\Phi}T^{a}_{b}
+ T^{a}_{{\rm m} b} \right) - T^{a}_{{\rm m} b;a}
= \frac{1}{3} \Phi \Phi_{,a} \left(R^{a}_{b} - \frac{1}{2} \delta^{a}_{b} R
\right) - T^{a}_{{\rm m} b;a} ~.
\eea
The assumption of the energy-momentum conservation for matter part
\bea
T^{a}_{{\rm m} b;a} &=& 0~,
\eea
would give the condition for the equivalence principle to be
fulfilled -- the assumption which was made in Brans-Dicke theory
(see e.g. \cite{weinberg}). However, one can make an assumption
that the matter is created in self-creation fashion
\cite{barber}, i.e., that
\bea
T^{a}_{{\rm m} b; a} &=& f(\Phi) \Box \Phi= f(\Phi) T_{\rm m}~,
\eea
which would make an agreement with a general requirement of
conformal invariance given by (\ref{tildelaws10}).

Because of the restriction to allow only traceless type of matter, for consistency of the
theory, we now admit a potential term into the action in order to get more degrees
of freedom while trying to keep conformal invariance.
Let us first introduce the potential for the
scalar field $\tilde{\Phi}$ (or $\Phi$) itself.
First, we try the common mass term
\bea
\label{massterm}
\tilde{V}(\tilde{\Phi}) &=& \frac{1}{2} \tilde{m}^2 \tilde{\Phi}^2~.
\eea
The action now looks as follows
\bea
\label{mass_conf1}
\tilde{\S} &=&
\frac{1}{2}\int~\d^4x~\stg\left[
  \frac{1}{6} \tilde{R}\tilde{\Phi}^2+
  \stackrel{\sim}{\partial}_{a}\tilde{\Phi}\stackrel{\sim}{\partial}^{a}\tilde{\Phi}
  +\tilde{m}^2\tilde{\Phi}^2
\right ] + \tilde{\S}_{\rm m}~,
\eea
which after using (\ref{conf_trafo}) and (\ref{PhitildetoPhi}) transforms into
\bea
\label{mass_conf2}
\S &=& \frac{1}{16\pi} \frac{1}{2}\int~\d^4x~\sg\left [\frac{1}{6} R \Phi^2 +
{\partial}_{a}\Phi {\partial}^{a}{\Phi}
  + m^2\Phi^2\right] + \S_{\rm m}~.
\eea
Thus, it is conformally invariant, if the mass scales as
\bea
\label{m_scal}
\tilde{m} &=& \Om^{-1}m~,
\eea
and the redefined mass term reads as (cf. Eq. (\ref{PhitildetoPhi}))
\bea
\label{Vphi}
V(\Phi)&=& \frac{1}{2} m^2 \Phi^2~.
\eea
Formally, the mass term contributes to the energy-momentum tensor
as follows
\bea
^V T_{ab} &=&  \frac{1}{2} g_{ab} m^2 \Phi^2~,\\
^V T &=& 2 m^2 \Phi^2~.
\eea
The appropriate field equations which result from (\ref{mass_conf1}) are
\bea
\label{eom4TV}
G_{ab} \equiv R_{ab}- \frac{1}{2}g_{ab}R
&=& \frac{6}{\Phi^2} \left[^{\Phi} T_{ab}  +
T^{\rm m}_{ab} \right] + \frac{1}{4} m^2 g_{ab}~,
\eea
from which we can immediately realize that the mass term plays the
role of the cosmological constant. The contraction of (\ref{eom4TV}) gives
\bea
\left( \Box - m^2 \Phi - \frac{1}{6} R \right) &=& 0~,
\eea
while varying (\ref{mass_conf2}) with respect to $\Phi$ we obtain
\bea
\left( \Box - m^2 \Phi - \frac{1}{6} R \right) &=& \frac{1}{\Phi} T_m~,
\eea
which again is consistent only for traceless matter
\bea
T_m &=& -\varrho + 3p =0~.
\eea
On the other hand, the application of Bianchi identity to
(\ref{eom4TV}) gives
\bea
\label{conslawmass}
^{\Phi}T^{a}_{~b;a} &=& 2\frac{\Phi_{,a}}{\Phi} \left[
^{\Phi} T^{a}_{~b}  + T^{a}_{{\rm m} b} \right]
- T^{a}_{{\rm m} b;a}~,
\eea
and the conservation law for ordinary matter
\bea
T^{a}_{{\rm m} b;a} = 0~
\eea
may or may not be imposed. It is interesting to note that the
contribution to (\ref{conslawmass}) from the mass term (\ref{Vphi}) has
been cancelled.

On the other hand, the self-interacting scalar field potential
\bea
\label{selfint}
\tilde{U}(\tilde{\Phi})=\frac{\tilde{\la}}{4}\tilde{\Phi}^4~,
\eea
with the coupling
constant $\tilde{\la}$ is self-conformally-invariant only in $D=4$
spacetime dimensions. In order to see this, we start with the
action with self-interaction potential which under conformal transformation
changes as
\bea
\S &=&
\frac{1}{2}\int~\d^Dx~\stg\left[
\frac{1}{6} \tilde{R}\tilde{\Phi}^2+
  \stackrel{\sim}{\partial}_{a}\tilde{\Phi}\stackrel{\sim}{\partial}^{a}\tilde{\Phi}
+\frac{\tilde{\la}}{4}\tilde{\Phi}^4 \right
]\nonumber  \\
&=&
\frac{1}{2}\int~\d^Dx~\sg\left [
\frac{1}{6} R \Phi^2 +
{\partial}_{a}\Phi {\partial}^{a}{\Phi}
+  \frac{\tilde{\la}}{4}\Om^{D-4}\Phi^4\right ] ~.
\eea
From this, we can immediately see that in order to get conformal
invariance one has to rescale the coupling constant as
\bea
\label{coupling_resc}
\tilde{\lambda} &=& \Om^{4-D} \lambda~,
\eea
which in $D=4$ dimensions gives a self-invariance. Now the self-interaction potential
reads as
\bea
U(\Phi)=\frac{\la}{4}\Phi^4~.
\eea
For a more general framework see e.g.~{\cite{polarski}}. This can
also be seen using similar arguments as in Eq. (\ref{eom4TV}).

It emerges that it is possible to maintain the conformal invariance of a fermion
field $\Psi(x)$ described by the Dirac equation, which applies to
spin-$1/2$ particles like quarks, electrons or protons.\nl
The Dirac equation can be derived in the framework of the
classical Lagrange field theory by varying the action
\bea
\label{diracaction}
S_D &=& \int~\d^Dx \sg \bar{\Psi}\left [ \imath \delsl - m \right ] \Psi~,
\eea
where $\delsl = \ga^{a}\del/\del x^{a}$ and $\ga^{a}$ are Dirac matrices, obeying
the anti-commutation relation
\bea
\left \{ \ga^{a},\ga^{b}\right \}_{+} &=& 2 g^{ab}~,
\eea
and
\bea
\ga^{a\dagger}&=&\ga^0\ga^{a}\ga^0~.
\eea
The convention we use is
\bea
\ga^{0} &=& \left (\begin{array}{cc} 1 & 0 \\ 0 & -1\end{array} \right
)~,{\rm and}~
\vec{\ga} = \left (\begin{array}{cc} 0 & \vec{\si} \\ -\vec{\si} & 0
  \end{array} \right )
\eea
where $\vec{\si}=(\si_1,\si_2,\si_3)$ are the Pauli matrices.
The adjoint field $\bar{\Psi}$ is defined by
\bea
\bar{\Psi} &=& \Psi^{\dagger}(x)\ga^0=\frac{\de \LL_D}{\de
  \Psi_{,0}}~.
\eea
The action (\ref{diracaction}) transforms under (\ref{conf_trafo}) as
\bea
\label{SDir}
\tilde{S}_D &=& \int~\d^Dx \stg
\bar{\tilde{\Psi}}\left(\imath\tilde{\delsl}-\tilde{m} \right )\tilde{\Psi}~.
\eea
and is conformally invariant provided that the fermion field transforms as
\cite{birell}
\bea
\bar{\tilde{\Psi}} &=& \Om^{\frac{1-D}{2}}\Psi~
\eea
and the masses scale as in (\ref{m_scal}).

For the vector boson (spin-1) field $V_{\mu}$ with strength
$F_{ab} = \partial_{a} V_{b} - \partial_{a} V_{b}$ action
\bea
\tilde{S}_V
&=& \int~\d^Dx \stg\left[-{1\over
    4}\tilde{F}_{ab}\tilde{F}^{ab}+{1\over2}\tilde{m}^2\tilde{V}_a
    \tilde{V}^{a}\right]~,
\eea
the application of (\ref{conf_trafo}) gives
\bea
S_V &=& \int~\d^Dx \sg\left[-{1\over
    4}F_{ab}F^{ab}+{1\over
    2}{\Om}^{D-2}\tilde{m}^2V_a V^{a}\right]~,
\label{Lem2}
\eea
which means that it can be conformally invariant \cite{bbpp}, provided the
masses scale according to (\ref{m_scal}), i.e.,
\bea
\tilde{\rm m}~&\rar&~\Om^{\frac{2-D}{2}}{\rm m}~ \nonumber.
\eea

It means that scaling of mass is an important possibility to
maintain the conformal invariance of the theory.

However, despite the idea is formally correct, it is a bit
striking in its physical context, for (\ref{m_scal}) makes the mass term in (\ref{mass_conf1})
coordinate-dependent. This is due to the fact that $\Om$ is coordinate-dependent. In other
words, the mass $m$ in (\ref{m_scal}) can be invariant and
coordinate-independent, while after the transformation the mass
$\tilde{m}$, as being equal to $\Om^{-1}m$, is coordinate-dependent.
Then, in this case, $\tilde{m}$ is not exactly what we mean by an invariant mass,
but it is rather a certain scalar field in the universe. This approach is acceptable
if we believe that the mass can really be a cosmic field
(cf. Refs. \cite{HN,bbpp,barber,canuto}). In the simple Friedmann
cosmology framework, it means that the mass is only the time-coordinate
dependent and it scales with the expansion of the universe
becoming effectively a cosmic scalar field \cite{bbpp}.

In a more
traditional approach \cite{shapiro_chi}, one tries to keep the covariant meaning of mass
after conformal transformation. This can be achieved, if one introduces a new scalar
field $\chi$ which transforms conformally in a way similar to (\ref{PhitildetoPhi}),
i.e., by
\be
\label{chi_trafo}
\tilde{\chi} = \Om^{-1} \chi .
\ee
In this case the mass of the scalar field $\Phi$ changes into
\be
\tilde{m}^2 = \frac{m^2}{M^2} \tilde{\chi}^2~,
\ee
with $M$ being a dimensional parameter. In such an approach, it is the field $\tilde{\chi}$
(sometimes called the cosmion \cite{wetterich87}), which takes a
coordinate-dependence rather than $\tilde{m}$. This saves the problem of a mass non-invariance
which is faced, if one assumes (\ref{m_scal}). This argument may also be applied
to a fermion and a vector mass terms given in (\ref{SDir}) and in (\ref{Lem2}), as well as
to a coupling constant rescaling in (\ref{coupling_resc}).

Let us finally mention that very general conformally invariant actions
based on gauged Wess-Zumino-Witten terms were studied in Ref.
\cite{anabalon}. Also, multidimensional $f(R)$ theory models which
explored conformal transformations were studied in Ref.
\cite{zhuk}.

\section{Conformal transformations as duality transformations in superstring theory}
\label{duality}
\setcounter{equation}{0}

It emerges that the conformal transformations under some special
conditions may behave like duality transformations in superstring
theory \cite{bekensteindual}. This can be shown easily by defining a conformal factor as
\bea
\label{conf_fact1}
\Omega(x) = e^{\omega(x)}~,
\eea
where $\omega(x)$ is a new scalar so that
\bea
\label{conf_fact2}
\tilde{g}_{ab}(x) &=& e^{2\omega(x)} g_{ab}(x)~,
\eea
and one can get much simpler rules of the transformation for the
geometric quantities of the Section \ref{conrel}. The Einstein
frame (with tildes) and Jordan frame (without tildes) quantities
can then be obtained by the simple duality transformation of the
form
\bea
\label{duality1}
\Omega \leftrightarrow \Omega^{-1}~~,
\eea
which is equivalent to
\bea
\label{duality2}
\omega \leftrightarrow - \omega~,
\eea
and corresponds to weak-strong coupling regime duality in superstring
theories if $\omega$ is a dilaton field \cite{polchinsky,superjim}.
Using (\ref{conf_fact1}) one has
\bea
\frac{\Omega_{,a}}{\Omega} &=& \omega_{,a}~\\
\frac{\Omega_{;ab}}{\Omega} &=& \omega_{;ab} +
\omega_{,a}\omega_{,b}~,\\
\frac{\Box{\Omega}}{\Omega} &=& \Box{\omega} +
\omega_{,a}\omega_{,b}~,
\eea
etc.
For the quantities calculated with respect to the Einstein frame
metric one has to replace $;$ with $\tilde{;}$ and $\Box$ with
$\stackrel{\sim}{\Box}$.

Below we give the rules of the conformal transformations of the
geometric quantities of Section \ref{conrel} with the conformal
factor given by (\ref{conf_fact1}). This proves especially useful while making
higher-order curvature calculations which involve complicated geometrical terms.

Since the rule of the conformal transformation (\ref{duality2})
now is dual-symmetric, one easily sees that the transition from the
quantities in a ``no-tilde'' frame into a ``tilde'' frame can be
just made by the simple replacement of the ``no-tilde'' quantities
to a ``tilde'' quantities and the replacement of $\omega$ into
$-\omega$, $\omega_{,a}$ into $-\omega_{,a}$, $\omega_{;ab}$ into
$-\omega_{\tilde{;}ab}$, $\Box{\omega}$ into $-\stackrel{\sim}{\Box} \om$
etc. Due to this, in order to avoid a simple repetition of the
conformal transformation rules in the text, we present only
``one-way'' transformations of the geometric quantities starting from the
``tilde'' frame and terminating in the ``no-tilde'' frame as follows further.

\noindent The Christoffel connection coefficients transform as
\bea
\label{connectionsom1}
\Ga^{c}_{ab}
&=&
\tilde{\Ga}^{c}_{ab}- \left( \tilde{g}^{c}_{a}
  \om_{,b} + \tilde{g}^{c}_{b} \om_{,a} -
  \tilde{g}_{ab}\tilde{g}^{cd}\om_{,d} \right)~,
  \hspace{0.5cm}\Ga^{b}_{ab}
= \tilde{\Ga}^{b}_{ab} - D \om_{,a}.
\eea
The Riemann tensor transforms as
\bea
\label{riemanntensorom2}
R^a_{~bcd} &=& \tilde{R}^a_{~bcd} - \left[\delta^a_d
\om_{\tilde{;}bc} - \delta^a_c \om_{\tilde{;}bd} + \tilde{g}_{bc}
\om^{\tilde{;}a}_{~\tilde{;}d}
- \tilde{g}_{bd} \om^{\tilde{;}a}_{~\tilde{;}c} \right] \\
&+&  \left[\delta^a_c
\om_{,b} \om_{,d} - \delta^a_d \om_{,b} \om_{,c} + \tilde{g}_{bd} \om^{,a}
\om_{,c} - \tilde{g}_{bc} \om^{,a} \om_{,d} \right]
+ \left[\delta^a_d \tilde{g}_{bc} - \delta^a_c \tilde{g}_{bd}
\right]\tilde{g}_{ef} \om^{,e} \om^{,f} ~~, \nonumber
\eea
the Ricci tensor transforms as
\bea
\label{riccitensorom2}
R_{ab} &=& \tilde{R}_{ab} + (D-2)\left[ \om_{,a}\om_{,b}
- \tilde{g}^{cd} \om_{,c}\om_{,d} \tilde{g}_{ab}\right ]
+ \left[(D-2)\om_{\tilde{;}ab}+
\tilde{g}_{ab} \stackrel{\sim}{\Box} \om  \right ]~,
\eea
the Ricci scalar transforms as
\bea
\label{ricciscalarom5}
R &=& e^{2\om} \left\{ \tilde{R} - (D-1) \left[
(D-2) \tilde{g}^{ab}
\om_{,a} \om_{,b} -2 \stackrel{\sim}{\Box}\om \right] \right\}~,
\eea
and the d'Alambertian transforms as
\bea
\Box\phi &=& e^{2\om} \left[ \stackrel{\sim}{\Box}\phi-
  (D-2)\tilde{g}^{ab} \om_{,a} \phi_{,b} \right ]~.
\eea
In fact, the obtained formulas (\ref{riemanntensorom2}),
(\ref{riccitensorom2}), and (\ref{ricciscalarom5}) agree with the
formulas (25), (26), and (27) given in Ref. \cite{shapiroconf}, provided
one puts $\omega = - \sigma$ in their notation.
The Einstein tensor transforms as
\bea
\label{Eintensorom2}
G_{ab} &=& \tilde{G}_{ab} + (D-2) \left[ \om_{,a} \om_{,b} + \frac{1}{2}
(D-3) \tilde{g}^{cd} \om_{,c}\om_{,d} \tilde{g}_{ab} \right]
+ (D-2) \left[\om_{\tilde{;}ab} - \tilde{g}_{ab} \stackrel{\sim}{\Box}\om
\right]~.
\eea
The curvature invariants transform as
\bea
\label{R2om1}
R^2 &=& e^{4\om} \left\{ \tilde{R}^2 + 4 (D-1) \stackrel{\sim}{\Box} \om
\left(\stackrel{\sim}{\Box} \om + R \right)
 \right.  \\ &+& \left.
(D-1)(D-2) \tilde{g}^{ab} \om_{,a} \om_{,b} \left[
(D-1)(D-2) \tilde{g}^{ef} \om_{,e} \om_{,f} - 4 (D-1) \stackrel{\sim}{\Box}\om
- 2 \tilde{R} \right] \right\}~,
\nonumber
\eea
\bea
\label{RabRabom1}
R_{ab}R^{ab} &=& e^{4\om} \left\{ \tilde{R}_{ab}\tilde{R}^{ab} + 2
\left[(D-2)\tilde{R}_{ab} \omega^{\tilde{;}ab} + \tilde{R} \stackrel{\sim}{\Box} \om \right] \right. \nonumber \\
&+& \left. (D-2)^2 \om_{\tilde{;}ab} \om^{\tilde{;}ab} + (3D -
4) \left( \stackrel{\sim}{\Box} \om \right)^2 + 2(D-2) \left[ \tilde{R}_{ab} \om^{,a} \om^{,b} -
\tilde{R} \tilde{g}^{ef} \om_{,e} \om_{,f} \right] \right. \nonumber \\
&-& \left. (D-2)^2 \om_{\tilde{;}ab} \om^{,a} \om^{,b} -
(D^2 + D - 3) \stackrel{\sim}{\Box} \om \tilde{g}^{ef} \om_{,e} \om_{,f} \right.
\nonumber \\ &+& \left. (D-1) (D^2 - 4D + 7) \left(\tilde{g}^{ef} \om_{,e} \om_{,f} \right)^2
\right\}~,
\eea
\bea
\label{Riem2om1}
R_{abcd} R^{abcd} &=& e^{4\om} \left\{\tilde{R}_{abcd} \tilde{R}^{abcd}
+ 8 \tilde{R}_{bc} \omega^{\tilde{;}bc} + 8 \tilde{R}_{bc} \om^{,b} \om^{,c}
- 4\tilde{R} \tilde{g}^{cd} \om_{,c} \om_{,d}
+ 4 \left( \stackrel{\sim}{\Box} \om \right)^2 \right. \nonumber \\
&+& \left. 4(D-2)\om_{\tilde{;}bc} \om^{\tilde{;}bc}
- 8 (D-2) \left[\stackrel{\sim}{\Box} \om \tilde{g}^{cd} \om_{,c} \om_{,d}
- \om_{\tilde{;}bc} \om^{,b} \om^{,c} \right] \right. \nonumber \\
&+& \left. 2 (D-1)(D-2) \left(\tilde{g}^{cd} \om_{,c} \om_{,d} \right)^2  \right\}~.
\eea
Again, the formulas (\ref{R2om1}), (\ref{RabRabom1}), and
(\ref{Riem2om1}) agree with the formulas (31), (30), and (29) of
Ref. \cite{shapiroconf}. Finally, the Gauss-Bonnet invariant transforms as
(compare the formula (35) of Ref. \cite{shapiroconf})
\bea
\label{omRGB}
R_{GB} &\equiv& R_{abcd} R^{abcd} - 4 R_{ab} R^{ab} + R^2 =
e^{4\om} \left\{ \tilde{R}_{GB} - 4(D-3) \left( 2
\tilde{R}_{ab} \omega^{\tilde{;}ab} - \tilde{R} \stackrel{\sim}{\Box} \om \right) \right.
\nonumber \\ &+& \left. 4 (D-3)(D-2) \left[\left(\stackrel{\sim}{\Box} \om \right)^2
- \om_{\tilde{;}ab} \om^{\tilde{;}ab} \right] -
2(D-3)(D-4) \tilde{R} \tilde{g}^{cd} \om_{,c} \om_{,d} \right. \nonumber \\
&-& \left. 8 (D-3) \tilde{R}_{ab} \om^{,a} \om^{,b} - 4(D-2)(D-3)^2 \stackrel{\sim}{\Box}
\om \tilde{g}^{cd} \om_{,c} \om_{,d}
- 8 (D-2)(D-3) \om_{\tilde{;}ab} \om^{,a} \om^{,b} \right. \nonumber \\
&+& \left. (D-1)(D-2)(D-3)(D-4) \left(\tilde{g}^{cd} \om_{,c} \om_{,d} \right)^2
\right\}~.
\eea

An interesting example of the duality-like symmetry was considered
in Ref. \cite{shapirodual} for a metric-dilaton model of the form
\bea
S = \int d^4 x \sqrt{-g} \left[A(\phi) g^{ab} \partial_a \phi
\partial_b \phi + B(\phi) R + C(\phi) \right]~,
\eea
where $A(\phi)$, $B(\phi)$, and $C(\phi)$ are the functions of the
dilaton which obey some constraints. A generalized variant of this
theory was studied in Ref. \cite{shapirodualgen}.

\section{Conclusion}

In this paper we discussed the rules of conformal transformations for geometric quantities
in general relativity such as connection coefficients, Riemann tensor, Ricci
tensor, Ricci scalar, Einstein tensor and the d'Alembertian operator in an arbitrary
spacetime dimension $D$. Since the conformal transformations are also used to
investigate higher-order gravity theories we also found the conformal transformation
rules for the curvature invariants $R^2$, $R_{ab}R^{ab}$, $R_{abcd}R^{abcd}$
and, as a consequence, for the Gauss-Bonnet invariant
in $D$ spacetime dimensions.

We devoted some effort in order to discuss precisely the conformal transformations of the
matter energy-momentum tensor and, in particular, the energy-momentum tensor composed of
the conformal factor $\Omega$. We showed that the conserved
energy-momentum tensor is not the same as the energy-momentum
tensor of the conformal factor and this is the reason why we may
say that the conformal transformation ``creates'' an extra matter term composed
of the conformal factor which enters the conservation law. In
other words, an empty Minkowski space after conformal
transformation may produce a non-zero energy-momentum tensor composed of
the conformal factor $\Omega$.

We also discussed how to construct the conformally invariant gravity. Its
simplest version is a special case of the scalar-tensor Brans-Dicke theory
- the one with the Brans-Dicke parameter $\omega = -3/2$. It can
be made conformally invariant due to the admission a non-minimal
coupling of the scalar field to gravity as well as due to the
admission of the appropriate kinetic term for a scalar field. We
have shown that the massive scalar field, self-interacting scalar
field, the Dirac field and the vector field theories can also be
made conformally invariant at the expense of the rescaling
appropriate fields and, in particular, of the mass scaling.

Finally, we presented already obtained rules of the conformal transformations
for the geometrical quantities in the fashion of the duality transformation
as in superstring theory. In such a case the transitions between conformal
frames can just be obtained by a simple change of the sign of the quantity $\om =
\ln{\Om}$, where $\Om$ is the conformal factor. We found these
rules the easiest of all possibilities.

We are aware of the fact that many studies of the problem of the
conformal transformations of the geometrical and physical quantities have 
been done so far. However, we decided to collect all of these rules
in one paper in order to give the reader a fairly comprehensive
collection of these transformations for future reference.

\section{Acknowledgments}

M.P.D. and J.G. acknowledge a partial support of the Polish Ministry of
Education and Science grant No N N202 1912 34 (years 2008-10)
while D.B.B. acknowledges a partial support of the grant No N N202 0953 33 (years 2007-10).


\end{document}